\documentclass[10pt,letterpaper,twocolumn]{article} 
\usepackage{ol2}
\usepackage[draft,implicit=false]{hyperref}
\usepackage{amsmath,amssymb,graphicx}

\begin{document}

\twocolumn[ 

\title{Pulse Evolution and Phase Sensitive Amplification in Silicon Waveguides}
\author{Y. Zhang,$^{*}$ C. Husko, J. Schr\"{o}der, and B. J. Eggleton}
\address{ Centre for Ultrahigh bandwidth Devices for Optical Systems (CUDOS),  Institute of Photonics and Optical Science (IPOS), School of Physics, University of Sydney, NSW 2006, Australia}
\email{y.zhang@physics.usyd.edu.au}
\begin{abstract}We provide an analytic solution for pulse propagation and phase sensitive amplification in silicon waveguides in the regime of strong two-photon absorption (TPA) and significant free carrier effects. Our analytic results clearly explain why and how the TPA and free carriers affect the signal gain. These observations are confirmed with numerical modelling and experimental results. 
\end{abstract} 

]
On-chip phase sensitive amplification (PSA) has been exploited for a variety of applications, including noiseless amplification in PPLN \cite{umeki2013line}, optical phase regeneration in silicon \cite{da2014phase}, and bandwith-tunable amplification in chalcogenide \cite{zhang2014chalwg}. Silicon is a compelling material for integrated PSA due to its compatibility with CMOS processing. A drawback of silicon is two-photon absorption (TPA) that significantly restricts the desirable Kerr effect through attenuation of the optical intensity and the generation of free carriers \cite{foster2006broad,yin2007impact} which cause free carrier absorption (FCA) and free-carrier dispersion (FCD). Although free carrier effects can be minimized by using low repetition rate pulses \cite{zhangphase} or PIN-junctions \cite{da2014phase}, detailed understanding of the limitations of TPA and free carriers on PSA and pulse-propagation is highly desired. 
The existing analytic descriptions of PSA consider the ideal (lossless) \cite{PSAinfiber2004} or linear loss \cite{zhang2014chalwg} cases,
while the solutions for pulse propagation in silicon only describe pulse evolution either without free carriers\cite{yin2007impact} or with small free carrier effects \cite{rukhlenko2009nonlinear}. In this work, we provide an analytic approach for four-wave mixing (FWM)-based 
silicon PSA in the highly nonlinear limit (with nonlinear phase shift up to 2 rad), including the effects of linear loss, TPA, FCA and FCD. The analytic method gives a clear insight into how TPA and free carriers modify the pulse evolution and the PSA gain. With our method, we predict the PSA gain under both pulsed and continuous wave pumping conditions. The results provide general guidelines for designing on-chip PSA in the presence of TPA and free carriers. 

For pump-degenerate PSA based on FWM, the pump wave ($\omega_{\rm p}$) is converted to the signal ($\omega_{\rm s}$) and idler ($\omega_{\rm i}$) waves given by the frequency relation $\omega_{\rm s}+\omega_{\rm i}=2\omega_{\rm p}$. When the three waves are seeded simultaneously the FWM process is phase sensitive. Thus we can control the signal amplification by detuning the phases of the waves. The FWM interaction in PSA is governed by a set of coupled mode equations \cite{lin2007nonlinear}. To obtain an approximate analytic solution, we make the following assumptions. First, the signal and idler powers are weak compared to the strong pump wave and we can ignore the signal and idler contribution to cross phase modulation and generation of free carriers. Second, it is reasonable to neglect the dispersion for an individual pulse of several picoseconds and longer since in a silicon chip the waveguide length is typically much shorter than the dispersion length $L_{\rm D}=T_0^2/|\beta_2|$, where $T_0$ is the pulse width and $\beta_2$ is the group index dispersion at pump wavelength. However, since the spectral separation between two adjacent waves is much larger than the spectral width of each wave, we must take into account the dispersion across the three waves. 

Before we solve the pump-degenerate FWM equations, we first consider the evolution of the pump wave, 
\begin{equation}
\frac {\partial E_{\rm p}}{\partial z}=(i\gamma_{r}-\frac {\gamma_{i}}{2})|E_{\rm p}|^2E_{\rm p}-\frac {\alpha}{2}E_{\rm p}-(\frac {\sigma}{2}+ik_0k_{\rm c})N_{\rm c}E_{\rm p},
\label{Eq:pump_only}
\end{equation}
where $E_{\rm p}(z,t)=E_{\rm p}$ is the slowly varying electric-field envelope of the pump wave, and $z$ is propagation distance. The nonlinearity is defined by $\gamma_{r}=k_0n_2/A_{\rm eff}$, where $n_2$ is the Kerr coefficient, $k_0= 2\pi/\lambda$, and $A_{\rm eff}$ is the effective mode area. The TPA coefficient $\alpha_{\rm TPA}$ is related to $\gamma_{i}=\alpha_{\rm TPA}/A_{\rm eff}$ and the linear loss is denoted by $\alpha$. The last term accounts for the absorption and dispersion of the free carriers with density of $N_{\rm c}(z,t)=N_{\rm c}$, where $\sigma$ is the FCA coefficient and $k_{\rm c}$ is the FCD coefficient \cite{yin2007impact}.

Substituting $E_{\rm p}=\sqrt {P} {\rm exp}\left(i \phi\right)$ into Eq. (\ref{Eq:pump_only}), we get the coupled differential equations for the output temporal power profile $P(z,t)$ and temporal phase profile $\phi(z,t)$
\begin{subequations}
\begin{align}
\partial P/\partial z&=-\alpha P-\gamma_i P^2-\sigma N_{\rm c}P, \label{Eq:der_opow} \\ 
\partial \phi/\partial z&=\gamma_r P-k_0k_{\rm c}N_{\rm c}.
\label{Eq:der_phs}
\end{align}
\label{Eq:der_opow_phs}
\end{subequations}
It is clear that TPA attenuates the wave power. For free carrier effects, the power profile is only affected by the FCA, while both the FCA (through $P$) and FCD modify the output phase. When there are no free carriers, i.e. $N_{\rm c}=0$, Eqs.~(\ref{Eq:der_opow}) and (\ref{Eq:der_phs}) can be solved to yield \cite{yin2007impact}
\begin{subequations}
\begin{align}
\widetilde{P}(z,t)&=\frac {P_0(t) {\rm e}^{-\alpha z}} {1+P_0(t)z_{\rm eff}\gamma_i}, \label{Eq:P_tpaonly}\\ 
\widetilde{\phi}(z,t)&=\gamma_r\gamma_i^{-1}{\rm ln}\left [1+\gamma_i P_0(t)z_{\rm eff}\right ], \label{Eq:phs_tpaonly}
\end{align}
\label{Eq:P_phs_taponly}
\end{subequations}
where $P_0(t)=\widetilde{P}(0,t)=P_0$ is the temporal power shape of the input wave, and $z_{\rm eff}=(1-{\rm e}^{-\alpha z})/\alpha$ is the effective distance. In the limit of low pump intensity, the effect of TPA is very small ($\gamma_i\approx 0$) and Eq.~(\ref{Eq:P_phs_taponly}) reduces to the case of linear loss only ($\widetilde{P}=P_0{\rm e}^{-\alpha z}$ and $\widetilde \phi=\gamma_r P_0z_{\rm eff}$), as in chalcogenide \cite{zhang2014chalwg}. 
 
In the more general case where free carriers play a significant role, we follow \cite{rukhlenko2009nonlinear} and assume the free carriers slightly modify the temporal profile of the pump wave. When the repetition rate ($R_{\rm e}$) of pulses is low ($R_{\rm e}T_0 \ll 1$), the power profile can be described by:
\begin{equation}
P(z,t)=\widetilde{P}(z,t)/\left[1+\eta(z,t)\right],
\label{Eq:P_hzt}
\end{equation}
where $\widetilde{P}(z,t)=\widetilde{P}$ is given in Eq.~(\ref{Eq:P_tpaonly}), and $\eta(z,t)=\eta$ is the FCA perturbation to the pulse shape. Substituting $P(z,t)$ into Eq.~(\ref{Eq:der_opow}), we obtain a differential equation $\frac {\partial \eta}{\partial z}=-\gamma_i \widetilde{P}\eta+\sigma N_{\rm c}(1+\eta)$. Rewriting the solution in the form:
\begin{equation}
\eta(z,t)=\widetilde{P} {\rm e}^{v(z,t)}\int^z_0 {\rm e}^{-v(z',t)s(z',t)}~dz', 
\label{Eq:h}
\end{equation}
and substituting Eq.~(\ref{Eq:h}) back into the differential equation, i.e. $\frac {\partial \eta}{\partial z}=-\gamma_i \widetilde{P}\eta+\sigma N_{\rm c}(1+\eta)$, we get $(\frac {\partial v}{\partial z}-\alpha)\eta+\widetilde{P}s=\sigma N_{\rm c}(1+\eta)$. This equation is true when the following relations are satisfied: $ \frac {\partial v}{\partial z}-\alpha=\widetilde{P}s=\sigma N_{\rm c}$. Therefore, $s$ and $v$ are found to be
\begin{equation}
s(z,t)= \frac{\sigma N_{\rm c}}{\widetilde{P}}, ~~
v(z,t)= \sigma\int^z_0 N_{\rm c}~dz'+\alpha z. 
\label{Eq:s,v}
\end{equation}

In Eq.~(\ref{Eq:s,v}), we still lack the expression for $N_{\rm c}$. The TPA-induced free carrier density $N_{\rm c}$ is governed by the rate equation $\frac {\partial N_{\rm c}}{\partial t}=\frac {\gamma_i}{2h\nu A_{\rm eff}}P^2-\frac{N_{\rm c}}{\tau_{\rm c}}$, where $h\nu$ is the energy of one photon and  $\tau_{\rm c}=1~\rm ns$ is the carrier lifetime in silicon \cite{yin2007impact}. When $R_{\rm e}T_0 \ll 1$, free carriers have enough time to recombine completely before the next pulse arrives and $N_{\rm c}$ can be solved at each propagation step 
\begin{equation}
N_{\rm c}(z,t)=\frac {\gamma_i}{2h\nu A_{\rm eff}}\int^t_{-\infty} {\rm e}^{-\frac{t-\tau}{\tau_{\rm c}}}P^2(z,\tau)~d\tau,
\label{Eq:Nc_int}
\end{equation}
where the integration tells us that the free carriers grow as the pulse passes through the waveguide. Now we have all the expressions to solve for $P$, Eq.~(\ref{Eq:P_hzt}). 

Since free carriers only slightly modify the pulse shape we set $P$ (free carrier perturbed) $\approx \widetilde{P}$ (TPA-only) in Eq.~(\ref{Eq:Nc_int}) for the initial condition. This yields
\begin{align}
\label{Eq:Ncbar}
\int^z_0 {N}_{\rm c}~dz' &\approx\frac {1}{2h\nu A_{\rm eff}}\int^t_{-\infty} {\rm e}^{-\frac{t-\tau}{\tau_{\rm c}}} \xi(z,\tau)~d\tau, \nonumber \\  
\xi(z,t)&=P_0-\widetilde{P}-\alpha \gamma_r^{-1}\widetilde {\phi}~.
\end{align}
Here $\widetilde P$ and $\widetilde \phi$ are given by Eq.~(\ref{Eq:P_phs_taponly}). Finally, we solve for the pulse shape $P$ by using Eq.~(\ref{Eq:Ncbar}) in Eqs.~(\ref{Eq:P_hzt})-(\ref{Eq:s,v}). 

Now that we have a solution for the pulse shape, we describe the phase profile. Integrating Eq.~(\ref{Eq:der_phs}) over propagation distance gives:
\begin{equation}
\phi(z,t)=\gamma_r \int^z_0P(z',t)~dz'-k_0k_{\rm c}\int^z_0 N_{\rm c}(z',t)~dz'.
\label{phs_out}
\end{equation} 
The phase shift is a combination of the Kerr effect and FCD. While the solution to the Kerr portion of this equation depends only on $P$, the free-carrier term is very sensitive to the form of $N_{\rm c}$. Importantly, here we include the effect of free-carriers on the pulse shape. Concretely, we use $P$ (perturbed) from Eq.~(\ref{Eq:P_hzt}) to calculate $N_{\rm c}$ in Eq.~(\ref{Eq:Nc_int}). Taking into account this FCA perturbation on $N_{\rm c}$ significantly widens the applicability of the present analytic solution compared to earlier work \cite{rukhlenko2009nonlinear} considering only TPA in the phase profile, i.e. using $\widetilde{P}$ (TPA-only) in Eq.~(\ref{Eq:Nc_int}). Later we will see that FCA significantly affects the output phase in our highly nonlinear limit. 

Together Eqs. (\ref{Eq:P_hzt}) and (\ref{phs_out}) provide an approximate solution of the output power and phase profiles of a pulse propagating along a silicon waveguide. It is easy to verify that, in the limit $\sigma \rightarrow 0$ (no FCA) and $k_{\rm c}\rightarrow 0$ (no FCD), we end up with the TPA only case, described in Eq. (\ref{Eq:P_phs_taponly}) \cite{rukhlenko2009nonlinear}.

\begin{figure}[htbp]
\centering
\includegraphics[trim=0cm 4cm 0cm 0cm, clip=true,width=1\columnwidth]{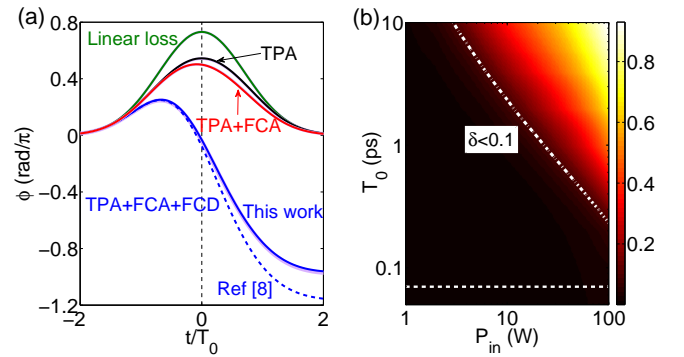}
\caption{(a) Output phases obtained from analytic solution (solid), modelling (light), and Ref \cite{rukhlenko2009nonlinear}. (b) Phase difference between analytic solution and modelling as a function of input powers and pulse widths. The whit dotted line indicates ${\rm \delta} = 0.1$ and dashed line corresponds $T_0=0.07 ~{\rm ps}$, where $L_{\rm D}$ is equal to waveguide length.}
\label{fig:com_malin}
\end{figure}

Since the pulse shape is less affected by free carriers than the phase profile, here we concentrate on the phase. To assess the accuracy of our analytic solution we compare with numerical simulations of the full pulse-propagation model \cite{yin2007impact}. 
Figure~\ref{fig:com_malin}(a) shows the evolution of the output 
phase in a $5~{\rm mm}$-length silicon waveguide with the different nonlinear effects. The parameters used in Fig.~\ref{fig:com_malin}(a) at $\lambda=1.55~\rm \mu m$ are $A_{\rm eff}=0.1~\rm{\mu m}^2$, $\alpha=1~\rm dB/cm$, $\alpha_{\rm TPA}=8\times 10^{-12}~\rm m/W$ \cite{soref1987electrooptical}, $n_2=6\times 10^{-18}~\rm m^2/W$, $\beta_2=-1~\rm ps^2/m$, $\sigma=1.45\times 10^{-21}~\rm m^2$ and $k_{\rm c}=3.4\times 10^{-27}~\rm m^3$ \cite{zhangphase}. An unchirped Gaussian pulse $P_0(t)=P_{\rm in}{\rm exp}(-t^2/T_0^2)$ is injected into the waveguide with $T_0=6~\rm ps$ and $P_{\rm in}=2~\rm W$. In Fig.~\ref{fig:com_malin}(a), the maximum phase is reduced in the presence of TPA compared to the case of linear loss only. 
We also notice free carriers induce pulse asymmetry in both the shape and phase. Physically this is explained by free carriers accumulating in the pulse tail as the pulse passes through the waveguide [see Eq.~(\ref{Eq:Nc_int})], causing absorption and dispersion. Notably, when the impact of FCD surpasses the Kerr effect at a certain time within the pulse, the total phase develops negative values \cite{yin2007impact}. In Fig.~\ref{fig:com_malin}(a), the dashed line is the phase profile calculated with the method proposed in \cite{rukhlenko2009nonlinear}, which shows the TPA-only approach loses its validity here at an intensity of $\rm 2~GW/cm^2$. In contrast, the excellent agreement with modelling confirms our improved analytic solution. This is attributed to the consideration of the FCA perturbation on the power shape, as mentioned before. 

To further quantify the accuracy of our solution, we calculate the difference between the output phase of the analytic solution ($\phi_{\rm a}$) and modelling ($\phi_{\rm m})$ as $ \rm \delta =\frac{ \int|\phi^2_{\rm a}-\phi^2_{\rm m}|}{\int \phi_{\rm m}^2}$. By integrating in the temporal domain from $-4T_0$ to $2T_0$, more than $99\%$ energy of the output pulse is included in our calculation. Figure~\ref{fig:com_malin}(b), shows $\delta$ as a function of the input peak power and the pulse width for a Gaussian pulse. As we can see $ \rm \delta$ increases at high power levels for all pulse widths. This can be explained by the ratio between the effects of FCA and TPA, $\frac{\rm FCA}{\rm TPA}=\frac {\sigma N_{\rm c}(P^2)}{\gamma_i P}$, which is approximately proportional to $P$. Therefore, higher powers imply stronger impact of free carriers on the pulse propagation. For the same reason, we can expect that an increase of the propagation length, the phase difference becomes more pronounced. At a fixed waveguide length, the shorter the pulse implies a larger acceptable input power. This is because for a given input power, longer pulses generate more free carriers than short pulses [see Eq.~(\ref{Eq:Nc_int})]. However, in the ultrashort pulse regime, we imagine this solution is to be further limited by dispersion. For example, when $T_0=0.2~\rm ps$ and $P_{\rm in}=2~\rm W$, dispersion affects the pulse significantly if waveguide length is longer than $10~\rm mm$ according to modelling results.

Next, we develop the analytic solution of PSA gain in silicon. We substitute the pump wave result $E_{\rm p}=\sqrt {P} {\rm exp}\left(i \phi\right)$ in Eqs.~(\ref{Eq:P_hzt}) and (\ref{phs_out}) as well as $N_{\rm c}$ in Eq.~(\ref{Eq:Nc_int}) into the following equation for the signal wave:
\begin{align}
\frac {\partial E_{\rm s}}{\partial z}&=2(i\gamma_r-\frac {\gamma_i}{2})|E_{\rm p}|^2E_{\rm s}-(\frac{\sigma}{2}+ik_0k_{\rm c})N_{\rm c}E_{\rm s} \nonumber \\ 
 & ~~~-\frac {\alpha}{2}E_{\rm s}+i\gamma_r E^*_{\rm i}E^2_{\rm p} {\rm exp}(-i \Delta \beta z),
 \label{Eq:signal}
\end{align}
where $E_{\rm s,i}$ are the electric-field envelopes of the signal and idler. The linear phase mismatch between input waves $\Delta \beta$ is given in the last FWM term. This can be expressed in terms of the total dispersion according to $\Delta\beta=\beta_{\rm 2}\Delta\omega^2+\frac{1}{12}\beta_{\rm 4}\Delta\omega^4$, where $\Delta\omega=|\omega_{\rm p}-\omega_{\rm s}|$, for dispersion orders up to $\beta_4$. We consider the case of identical signal and idler ($E_{\rm s}=E_{\rm i}$). Rewriting Eq.~(\ref{Eq:signal}) in form of $\frac{\partial E_{\rm s}}{\partial z} =i p_{\rm k} E_{\rm s}+ir_{\rm k}{\rm exp}\left(i\int^z_0q_{\rm k} dz'\right)$, we follow the interaction between four waves \cite{marhic2008fiber}. When the pump power varies slowly along propagation ($\partial r_k/\partial z\approx 0$), the complex gain ($\Gamma=E_{\rm s}/E_{\rm s,in}$) including the output amplitude and phase normalized to the input signal ($E_{\rm s,in}$) is given by
\begin{align} 
\Gamma(t)&=\sinh(gz)\left[\coth(gz)+i(k+\alpha_t)(2g)^{-1}\right]\exp(i\theta) \nonumber \\
&~~+i\sinh(gz)\gamma_r Pg^{-1} {\rm exp}\left(-i\theta \right),
\label{Eq:gain}
\end{align} 
where $\theta$ is the phase detuning between the signal, idler and pump, $\alpha_t=\alpha+2P\gamma_i+\sigma N_{\rm c}$ is the total loss including linear loss, TPA and FCA. The parametric gain parameter ($g$) is given by $g^2=(\gamma_r P)^2-\left(\frac{k+\alpha_t}{2}\right)^2$ with phase mismatch $k=2\gamma_r P+\Delta\beta$. It is interesting that FCD does not affect the phase matching, which agrees with the conclusion for FWM \cite{lin2007nonlinear}. The cancellation of the FCD term is due to the three waves experiencing exactly the same free carrier effects. At perfect phase matching $k=0$ and $\widetilde{g}^2=(\gamma_r P)^2-\left(\frac{\alpha_t}{2}\right)^2$, the signal gain of the intensity (${\rm G}=|\Gamma|^2$) is found to be
\begin{align} 
G(t)&= {\rm sinh}^2(\widetilde{g} z)\gamma_r P \widetilde{g}^{-1}\left[2\coth(\widetilde{g} z)-\alpha_{\rm t}\widetilde{g}^{-1}\right]\sin(2\theta) \nonumber \\ 
&+\sinh^2(\widetilde{g} z)\left[\left(\coth(\widetilde{g} z)-\alpha_{\rm t}(2\widetilde{g})^{-1}\right)^2+(\gamma_r P\widetilde{g}^{-1})^2\right],
\label{Eq:gain_k_0}
\end{align}
where the common phase term is ignored as it only adds $\pi/4$ phase to $\theta$ \cite{zhang2014chalwg}. The first term with $\rm {sin} (2\theta)$ in Eq.~(\ref{Eq:gain_k_0}) clearly highlights that the PSA gain is a sinusoidal function of $2\theta$. However, the amplitude [${\rm sinh}^2(\widetilde{g} z)\gamma_r P \widetilde{g}^{-1}(2\coth(\widetilde{g} z)-\alpha_{\rm t}\widetilde{g}^{-1})$] of this function is reduced as TPA and FCA introduce additional losses to $\alpha_{\rm t}$. Due to the same reason, the reference position of the gain determined by the second term is shifted down vertically. 

\begin{figure}[htbp]
\centering
\includegraphics[trim=0cm 4cm 0cm 0cm, clip=true,width=1\columnwidth]{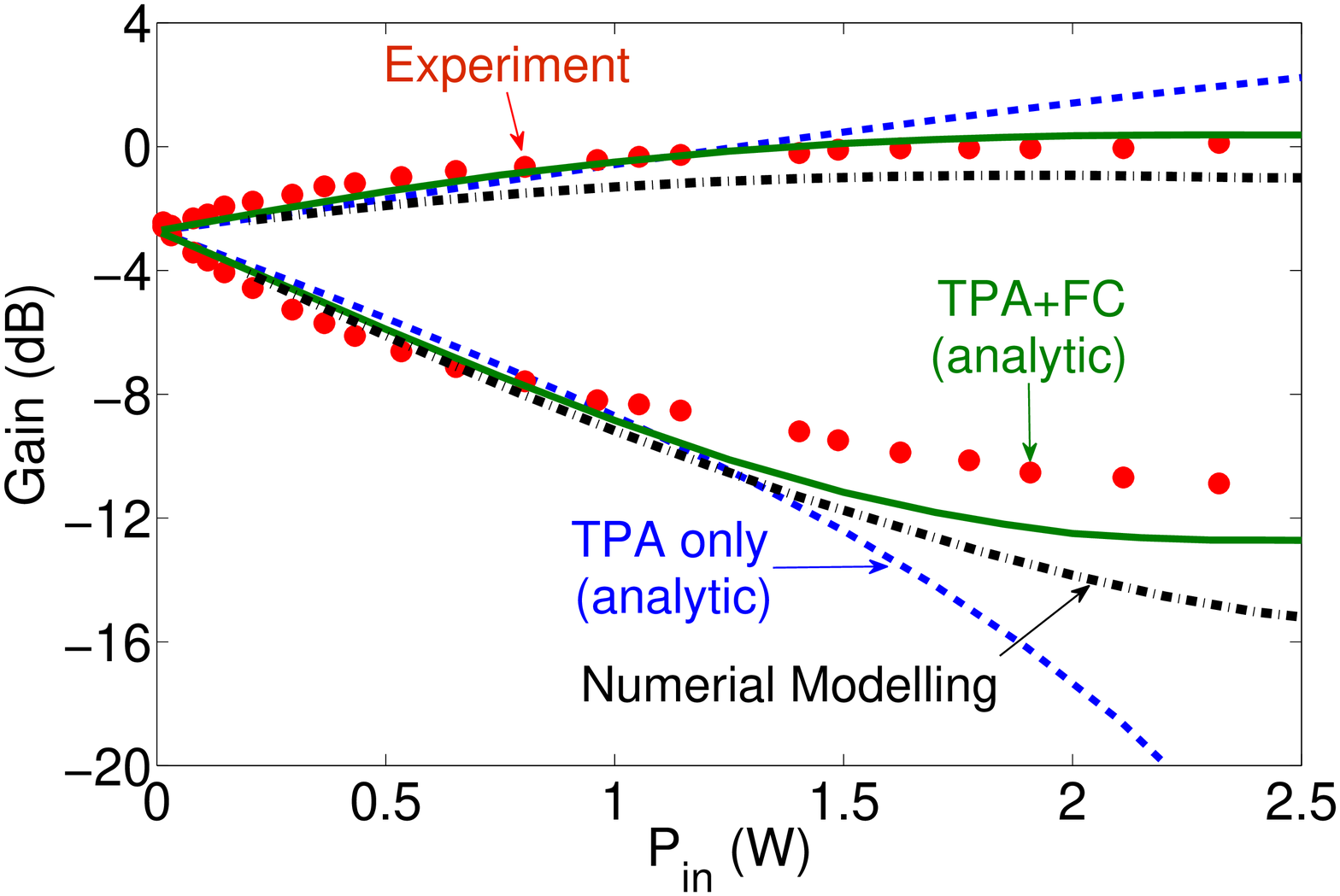}
\caption{PSA gain obtained from modelling, analytic solution and experiments vs input power.}
\label{fig:ER_pow_L}
\end{figure}

We compare our analytic solution with and without free carrier effects with experimental results from \cite{zhangphase}. Note we take into account slow light enhancement in photonic crystal waveguides \cite{zhangphase}. Figure~\ref{fig:ER_pow_L} shows the maximum and minimum signal gains obtained from the analytic solution presented here compared with experimental results \cite{zhangphase} as a function of the input peak power. In general, the analytic solution shows excellent agreement with modelling and experimental results. At low pump powers ($P_{\rm in}\textless 1~\rm W$) linear loss and TPA are dominant. As the pump power increases above $P_{\rm in}\textgreater 1.5~\rm W$, the free carriers (FC) start to play an important role. The comparison between the cases with and without FC demonstrates that FC effects shift maximum and minimum gain downward, as expected from Eq.~(\ref{Eq:gain_k_0}). In Fig.~(\ref{fig:ER_pow_L}), the free carriers saturate the maximum gain at high power levels compared to no observed saturation with TPA only. The free carrier induced saturation may also explain the gain saturation of FWM-based parametric amplification observed with picosecond pump pulses in silicon waveguides \cite{foster2006broad}. 

\begin{figure}[htbp]
\centering
\includegraphics[trim=0cm 5cm 0cm 0cm, clip=true,width=1\columnwidth]{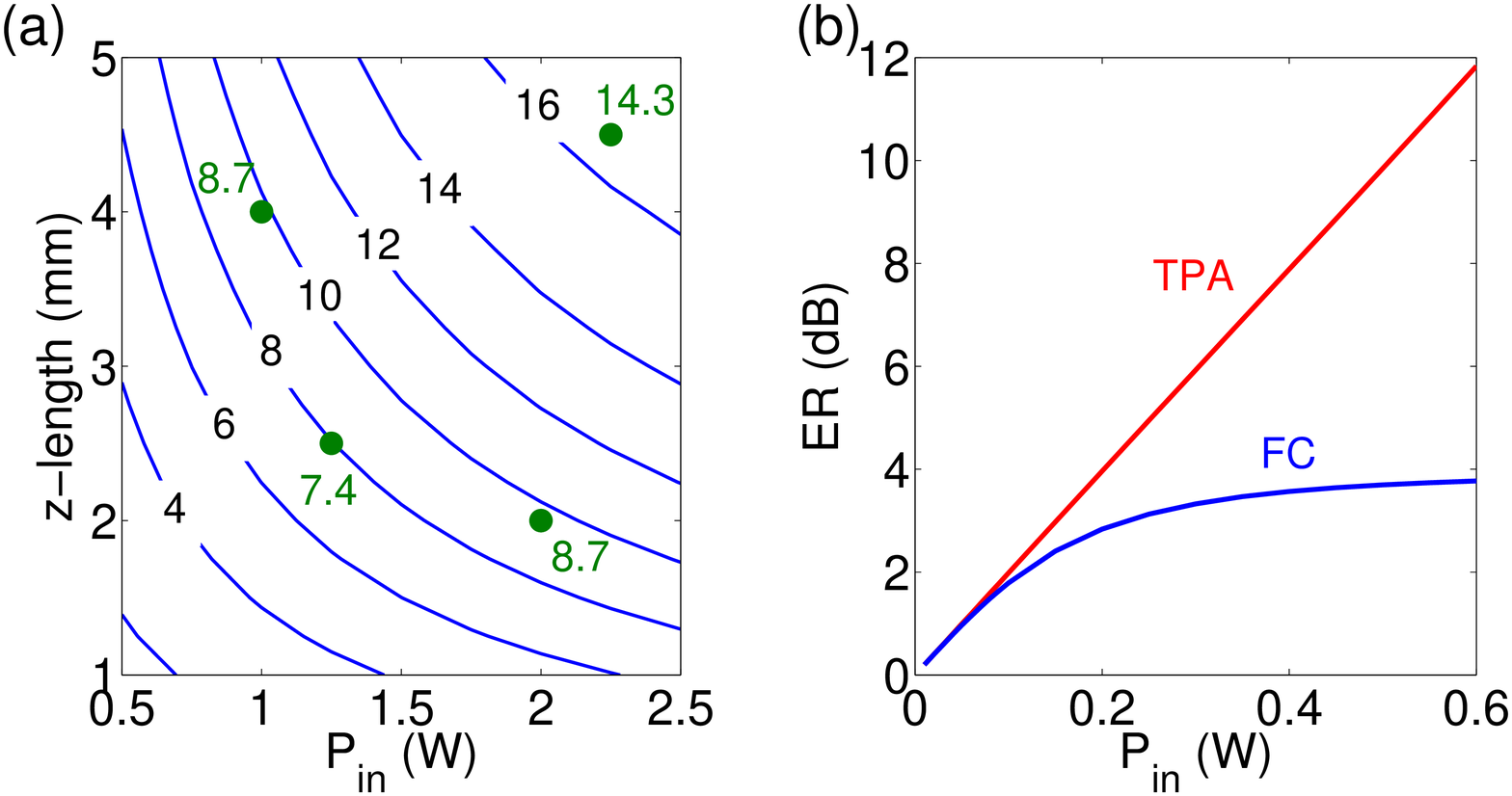}
\caption{(a) ER as a function of waveguide length and input power, where dots are calculated from modelling. (b) ER vs input power with and without free carriers.}
\label{fig:ER_tpa}
\end{figure}

Figure~\ref{fig:ER_tpa}(a) shows the extinction ratio (ER) of PSA as a function of waveguide length and peak power calculated with Eq.~(\ref{Eq:gain}). The ER is an important parameter to characterize the PSA performance in applications, for example phase regeneration \cite{da2014phase}. The three Gaussian pulses have $T_{0}=6~\rm ps$ and $\Delta\omega=7.7~\rm Trad/s$. The ER is the difference between the maximum gain and the minimum gain. The intensity gain is defined as the integration of the amplified signal in the temporal domain [$\int G(t)$]. 
In general, the ER increases at higher power levels with longer propagation distance. This gives flexibility designing a PSA when either waveguide length or available pump power is limited. For example, an ER of $8~\rm dB$ is found at either $1.3~\rm mm$-length waveguide with $P_{\rm in}=2.5~\rm W$ or $5~\rm mm$-length waveguide with $P_{\rm in}=0.6~\rm W$. However, this flexibility gradually reduces for higher ER. The maximum ER of $16~\rm dB$ that can be obtained is at the waveguide length around $4.5~\rm mm$ with a pump power round $2~\rm W$. 

Thus far we have discussed only the scenario of low repetition rate pulses ($R_{\rm e}T_0 \ll 1$). Our analytic method is also applicable in the high repetition rate regime ($R_{\rm e}T_0 \gg 1$), for example in optical communications. The PSA gain in Eq.~(\ref{Eq:gain}) still holds, however, a new formula for the output pump power is needed because of the strong FCA. Due to free carrier accumulation, $N_{\rm c}$ is estimated from the steady state of the rate equation, i.e. $\partial N_{\rm c}/\partial t=0$. Since linear loss and FCA are dominant, TPA can be ignored in Eq.~(\ref{Eq:der_opow}). The output power is then obtained following \cite{husko2011effect}
\begin{equation}
P=P_0 {\rm e}^{-\alpha z}\left[1+(\gamma_i\tau_c \sigma P_0^2 z_{\rm 2eff})/(h \nu A_{\rm eff})\right]^{-1/2},
\label{Eq:P_cw}
\end{equation}
where $z_{\rm 2eff}=(1-{\rm e}^{-2\alpha z})/(2\alpha)$. Figure~\ref{fig:ER_tpa}(b) shows the predicted ER with free carriers [Eq.~(\ref{Eq:P_cw})] and without free carriers [Eq.~(\ref{Eq:P_tpaonly})] for a CW pump. As expected, the free carriers saturate the ER very quickly compared to TPA-only case. From this analysis it is clear a PIN junction is required to minimize the FCA to obtain a desirable ER for optical regeneration \cite{da2014phase}.

In conclusion, we developed an analytic solution of pulse evolution and PSA in silicon. The method is crucial for understanding the effect of TPA and free carriers on pulse propagation and PSA gain in the highly nonlinear regime (with nonlinear phase up to 2 rad). This analytic method can be extended to materials limited by three-photon absorption such as silicon beyond 2.2 $\mu$m \cite{liu2010mid} or wide-gap materials \cite{husko2011effect} taking into account the appropriate equations. 

This work was supported by the Australian Research Councils, Laureate Fellowship (FL120100029), Center of Excellence CUDOS (CE110001018) and Discovery Early Career Researcher (DE120101329, DE120102069) schemes. 


\newpage
\begin{figure}[htbp]
\centering
\includegraphics[trim=0cm 0cm 0cm 0cm, clip=true,width=2\columnwidth]{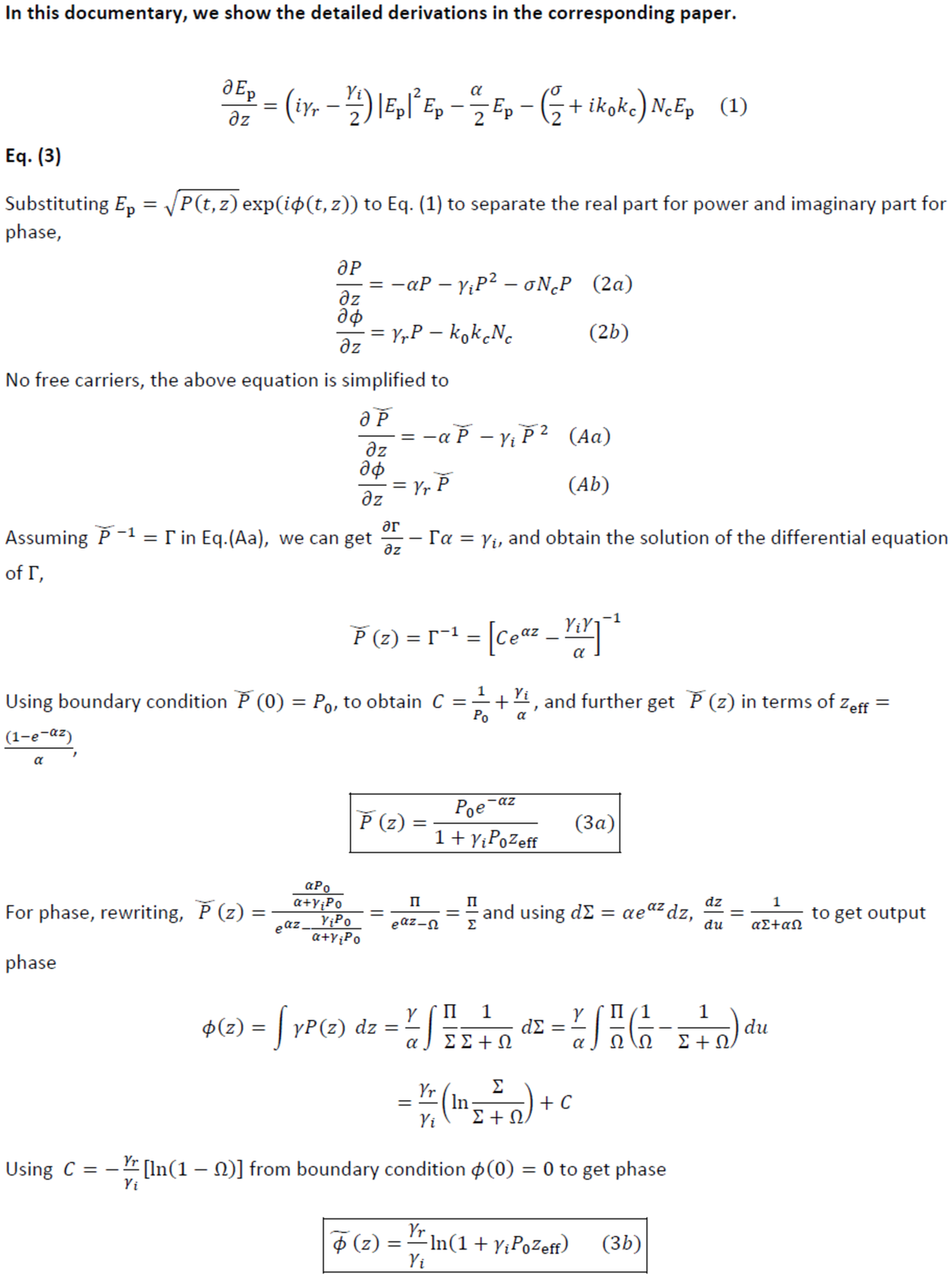}
\end{figure}
~~~~~~~~~~~~~~
\newpage

\begin{figure}[htbp]
\centering
\includegraphics[trim=0cm 0cm 0cm 0cm, clip=true,width=2\columnwidth]{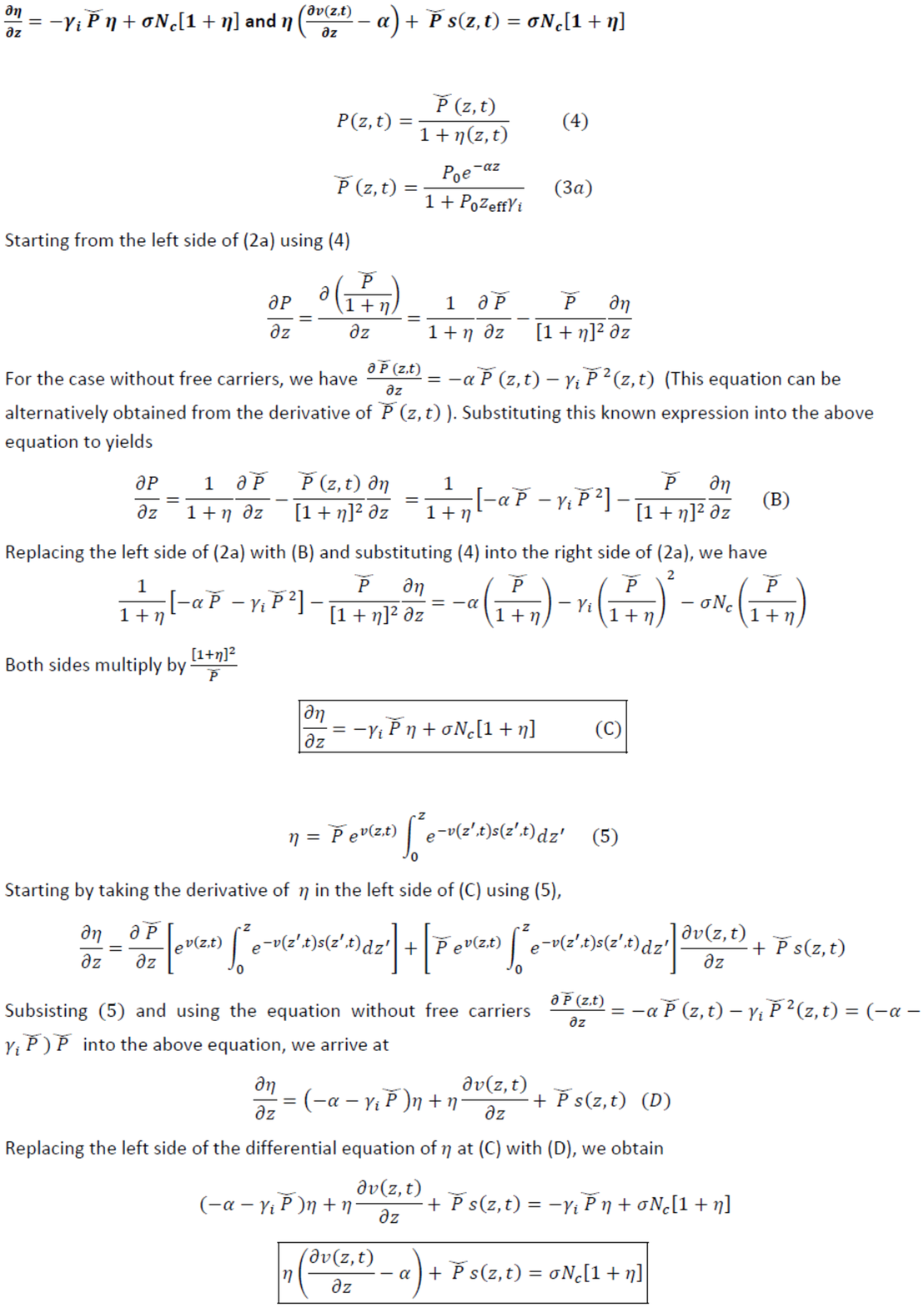}
\end{figure}
~~~~~~~~~~~~~~
\newpage

\begin{figure}[htbp]
\centering
\includegraphics[trim=0cm 0cm 0cm 0cm, clip=true,width=2\columnwidth]{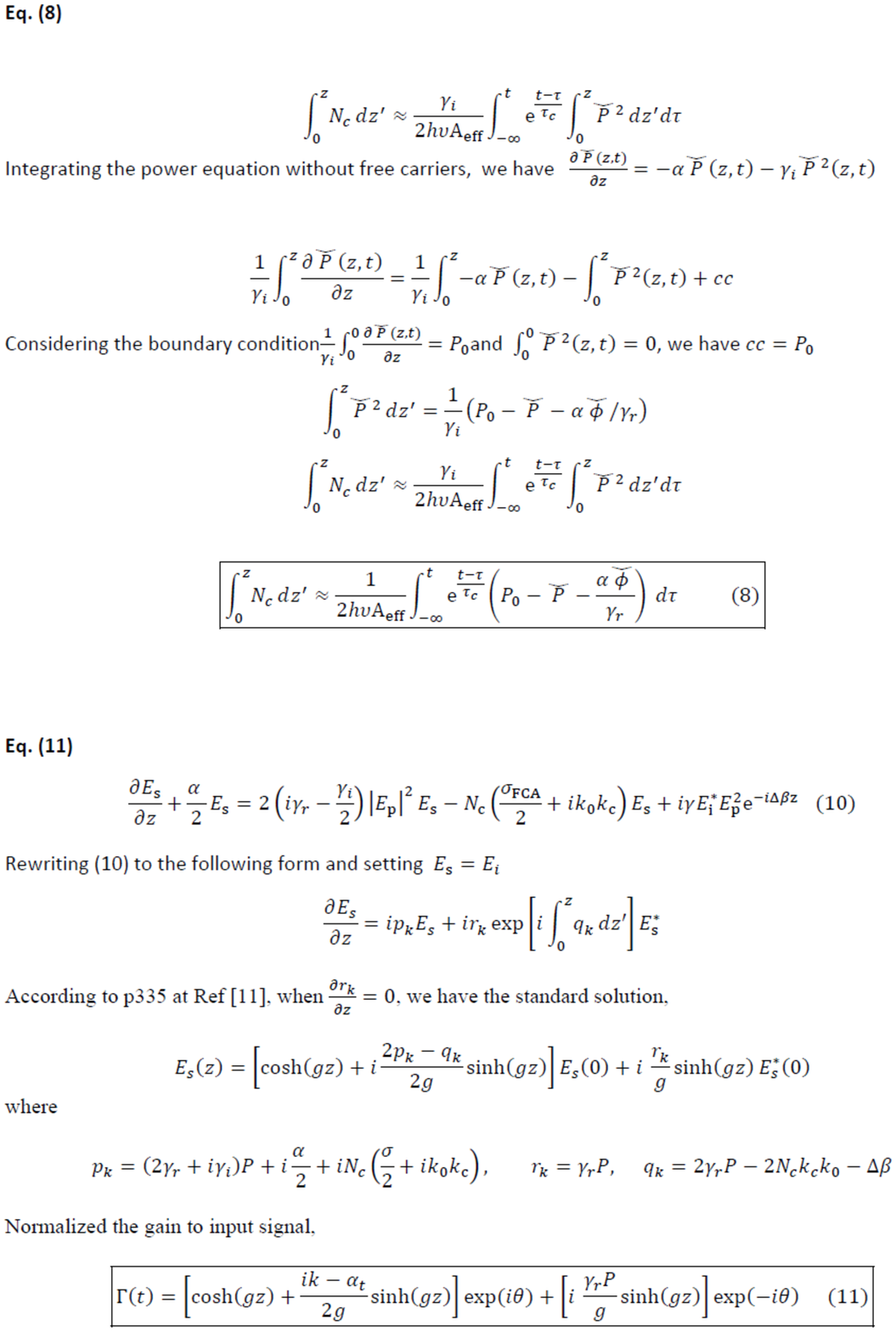}
\end{figure}
~~~~~~~~~~~~~~
\newpage
~~~~~~~~~~~~~~~~
\newpage
\begin{figure}[htbp]
\centering
\includegraphics[trim=0cm 0cm 0cm 0cm, clip=true,width=2\columnwidth]{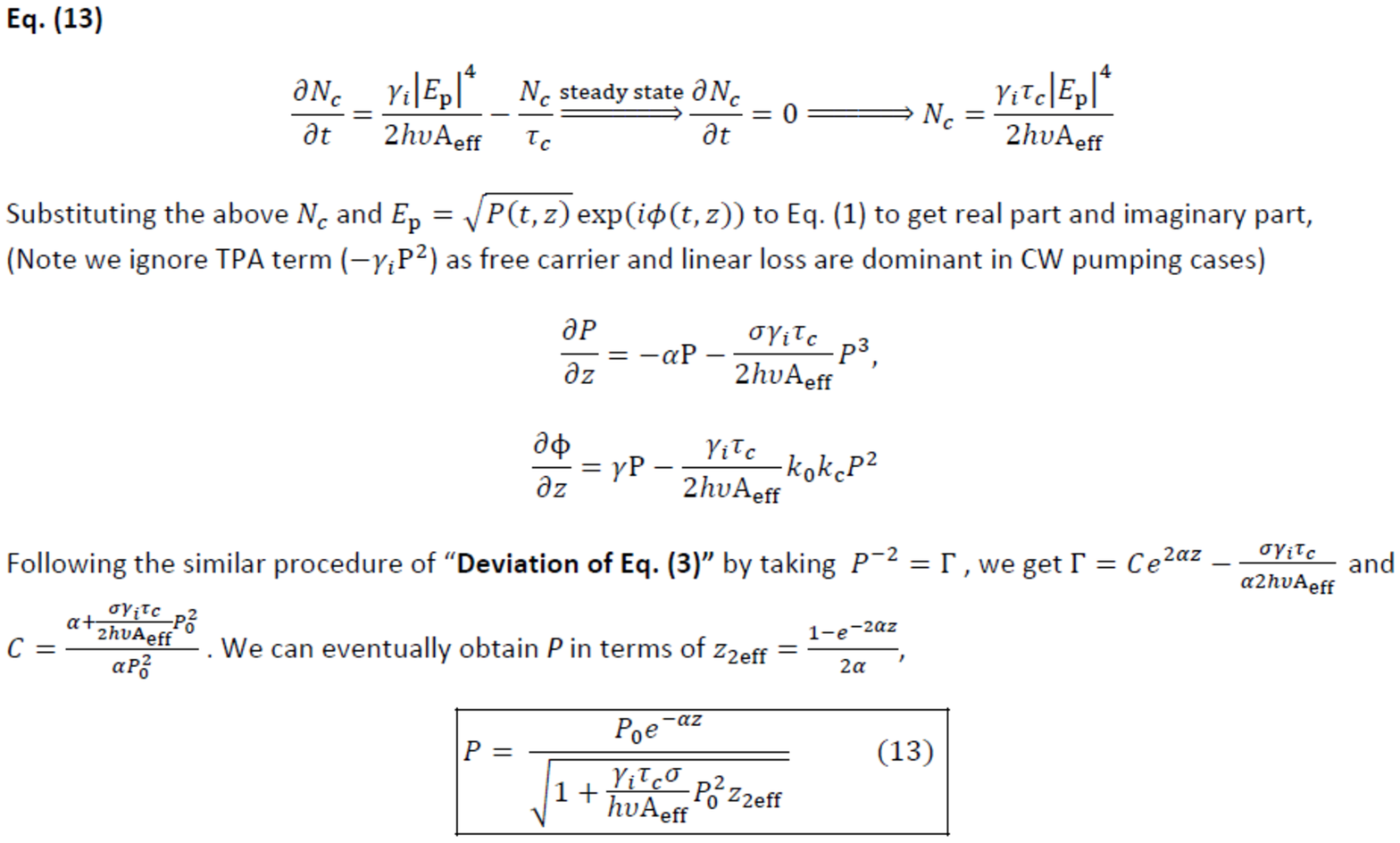}
\end{figure}
\end{document}